\documentclass[10pt,letterpaper]{article}

 \newcommand{\LN}{LiNbO$_3$~}

\usepackage{opex3}
\usepackage{graphicx}
\usepackage{amsmath}
\usepackage{verbatim} 
\usepackage{color}
\usepackage{url}
\usepackage{cite}

\begin{document}

\title{ Design of nanobeam photonic crystal resonators for a silicon-on-lithium-niobate platform}
\author{Jeremy D. Witmer$^{1}$, Jeff T. Hill and Amir H. Safavi-Naeini$^{2}$}

\affil{Ginzton Laboratory, Stanford University, Stanford, California 94305, USA\\
$^{1}$jwitmer@stanford.edu\\
$^{2}$safavi@stanford.edu
}



\begin{abstract}
We outline the design for a photonic crystal resonator made in a hybrid Silicon/Lithium Niobate material system. Using the index contrast between silicon and lithium niobate, it is possible to guide and confine photonic resonances in a thin film of silicon bonded on top of lithium niobate. Quality factors greater than $10^6$ at optical wavelength scale mode volumes are achievable. We show that patterning electrodes on such a system can yield an electro-optic coupling rate of 0.6 GHz/V (4 pm/V). 
\end{abstract}











\section{Introduction}

Resonant silicon photonic devices, though capable of achieving very large quality factors with wavelength-scale mode volumes, suffer from the lack of large nonlinear optical ($\chi^{(2)}$), piezoelectric, and electro-optic coefficients. This has led to recent research into other materials such as GaP, GaAs, and AlN which can outperform silicon in many respects~\cite{Shambat2010,Xiong2012,Tadesse2014,Balram2015}. Nonetheless, the appeal of large-scale integration with silicon photonics (and its rapidly developing toolkit as well as several foundries) leads us to consider optically nonlinear materials that can be heterogeneously integrated with high quality silicon passive structures. Lithium niobate (LN) is a technologically important ferroelectric which has some of the largest nonlinear optical coefficients found in a  bulk material, and can be obtained in ample quantity and high quality as a result of the demand from the telecommunications market. Unfortunately, it is far less amenable to microfabrication techniques than silicon, and processes such as dry etching high-quality wavelength scale optical structures are difficult and non-standard. Nonetheless, ion sliced thin films of LN have been developed in the last few years~\cite{Rabiei2004,Sulser2009,Poberaj2012,Lu2012,Rabiei2013} to facilitate among other things nanophotonic fabrication, and more recently, high quality chip-scale optical resonators have been demonstrated in these materials~\cite{Diziain2013,Chen2013,Lin2015,Wang2014,Wang2015}. 

In a recent work on mid-infrared modulators, thin-film \emph{silicon} was wafer-bonded to LN~\cite{Chiles2014}. Very recently, CMOS-compatible integration of SOI photonic waveguides with LN has also been demonstrated~\cite{Weigel2015}. The major advantage of this method is the ease of integration with silicon photonics. Fabrication of the silicon device layer can be accomplished at a variety of different foundries, and the back-end processing involves only a single bonding and back-etch step. 

A  question we address in this work is whether it is possible to effectively confine light inside LN by only patterning a top bonded silicon layer. Effectiveness is this context is judged by confinement and optical resonator decay rate, and is encapsulated in the ratio of the quality factor to optical mode volume, $Q/V$. We design a device which shows record  large $Q/V$ in the Si/LN material system and theoretically outline its performance characteristics and possible applications.

We start by introducing the Silicon/Lithium Niobate (Si/LN) platform in Section~\ref{sec:siln}, and by describing some of the fabrication flow involved. In Section~\ref{sec:cavity_design_sim} we introduce the design steps involved in generating a cavity on this platform. We calculate the expected electro-optic coupling coefficient obtained in Section~\ref{sec:eo}.







\section{Si/LN Platform Fabrication Flow}\label{sec:siln}

\begin{figure}
    \centering
\includegraphics[width=5in]{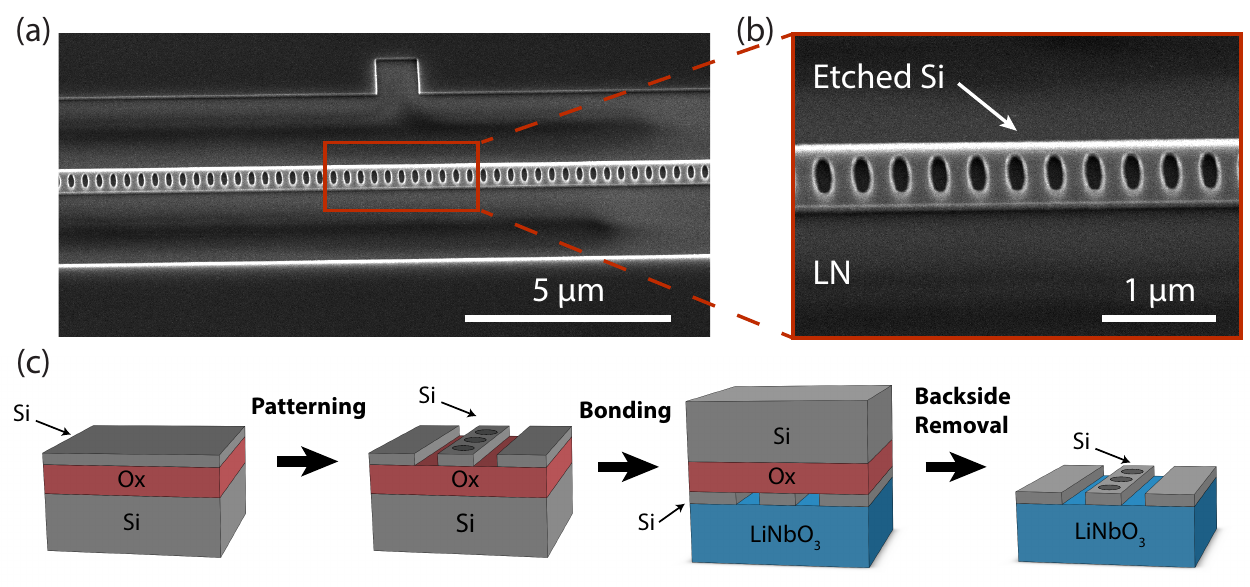}
    \caption{(a) An SEM image of a fabricated nanobeam cavity in silicon-on-insulator bonded to LN. An enlarged image (b) shows the elliptical holes which generate the photonic band gap. (c) shows the fabrication steps for creating Si/LN platform.  First, the crystalline Si device layer can be patterned using standard electron beam lithography techniques and silicon etching.  The second step is room temperature bonding of the SOI wafer to the LN substrate, enabled by a surface-activating plasma treatment.  Finally, the Si backside is removed (using a combination of mechanical polishing and wet or dry etching), followed by an HF dip to remove the oxide layer. }
    \label{fig:fab_steps}
\end{figure}

One promising platform which has recently been demonstrated is the so-called Si/LN platform.  This material system consists of a thin layer of high quality crystalline silicon from an SOI wafer transferred onto a bulk \LN wafer and is described in detail in \cite{Chiles2014}.  The fabrication steps that we pursue for this process are shown in Fig. \ref{fig:fab_steps} and  outlined below.

First, the silicon top layer is patterned using either e-beam or DUV lithography and dry etching. This process is CMOS compatible and can be done at a foundry, though the highest quality factor photonic resonators are typically fabricated with e-beam lithography and optimized etches. The SOI wafer and the LN wafer are then bonded together using a room-temperature direct bonding process (as described, eg. in \cite{Tulli2011, Takagi2001}).  A key limitation in the bonding process is the large thermal expansion coefficient mismatch between Si and LN ($2.6\times10^{-6}~\text{K}^{-1}$ for Si compared to $15.7\times 10^{-6}~\text{K}^{-1}$ for LN along the x and y axes \cite{Nikogosyan2005}), which limits the temperature at which the bonding process can occur.  The need for a high temperature step is averted by using an O$_2$ or Ar plasma treatment to activate the surface.  This treatment introduces damage to the surface and creates dangling bonds, resulting in hydrophilic surfaces with a high surface energy \cite{Tulli2011}.  After surface activation a strong bond can be achieved by applying only a minimal amount of pressure. The bulk Si can be removed using a combination of mechanical lapping and either a wet chemical etch or a dry plasma etch.  The exposed SiO$_2$ can then be removed using a simple HF dip. It should be stressed that a key advantage of this platform is that it avoids the need to etch the LN.  Figure \ref{fig:fab_steps}(a) shows an example 1D photonic crystal, fabricated in SOI using e-beam lithography and silicon etching, after being bonded to LN.  

It should be mentioned that it is also possible to pattern the Si device layer after bonding the two wafers together and removing the SOI backside, which is the method used in \cite{Chiles2014}.  The main advantage of patterning the Si device layer \textit{before} bonding is that it allows us to use silicon photonics processes (eg. etch tools) without the danger of contaminating sensitive CMOS silicon processing tools with lithium (or niobium). In fact, groups without access to lithography or silicon etching can simply bond foundry-fabricated silicon photonics devices onto LN.  It is also worth noting that instead of bulk LN wafers, so-called ``lithium-niobate-on-insulator'' wafers which have a thin LN film can also be used without significantly changing the fabrication flow (as in \cite{Weigel2015}).

\section{Cavity Design and Simulation}\label{sec:cavity_design_sim}

Photonic crystal resonators in quasi-1D and 2D systems are implemented by fabricating a periodic array of holes into an optically thin beam or slab. The periodic variation of the dielectric constant leads to a photonic bandgap for index-confined waves, which can then be used to confine light in all dimensions through the introduction of a defect. In silicon, quality factors on the order of $10^6$ and approaching $10^7$ have been demonstrated for both 1D and 2D structures of this type \cite{Akahane2003a, Sekoguchi2014, Deotare2009}. 

We follow a popular recipe for the creation of a high-$Q$ localized mode in a photonic crystal slab~\cite{Song2005,md2008ultra}. Firstly, we design a one-dimensional unit cell that has a photonic band gap at the resonant frequency of interest. This will form the ``mirror regions'' surrounding the cavity. Secondly, a central defect is introduced between the two mirror regions. This defect is generated by modifying the properties of the unit cell so that it supports modes with frequencies within the mirror region bandgap. The defect is introduced as ``smoothly'' as possible to prevent scattering of light into radiation modes, while also allowing for tight confinement of the light. Such cavities have been designed for a variety of suspended membrane materials such as silicon~\cite{Deotare2009}, GaAs~\cite{Buckley2014a}, and even diamond~\cite{Burek2014}. Non-suspended structures, i.e., photonic crystals made from silicon on top of glass, have also been demonstrated~\cite{md2008ultra}. In our case, since the silicon is surrounded on one side by air and on the other by LN, which has a fairly high index of $n \approx 2.2$, care must be taken to ensure that the nanobeam modes that would be bound for a beam in air or on glass do not leak into the LN substrate.  


Several considerations go into designing the unit cell. In comparison to suspended silicon photonic crystals, the presence of the LN substrate causes a red-shift of the Bloch-waves which can be countered by reducing the lattice spacing. The presence of the LN also raises the lower limit of how small the effective index can be, since the nearby medium (LN) has an index of approximately $2.2$. This leads to a higher filling factor of silicon in the photonic structure or, equivalently, smaller holes. Finally, an important practical constraint on the unit cell geometry is that it should be fabricable using standard e-beam lithography techniques and extendable to DUV photolighography for foundry processes.  We therefore adopted a minimum feature size of 75 nm as a design rule.

\begin{figure}[h]
\begin{center}
\includegraphics[width=5in]{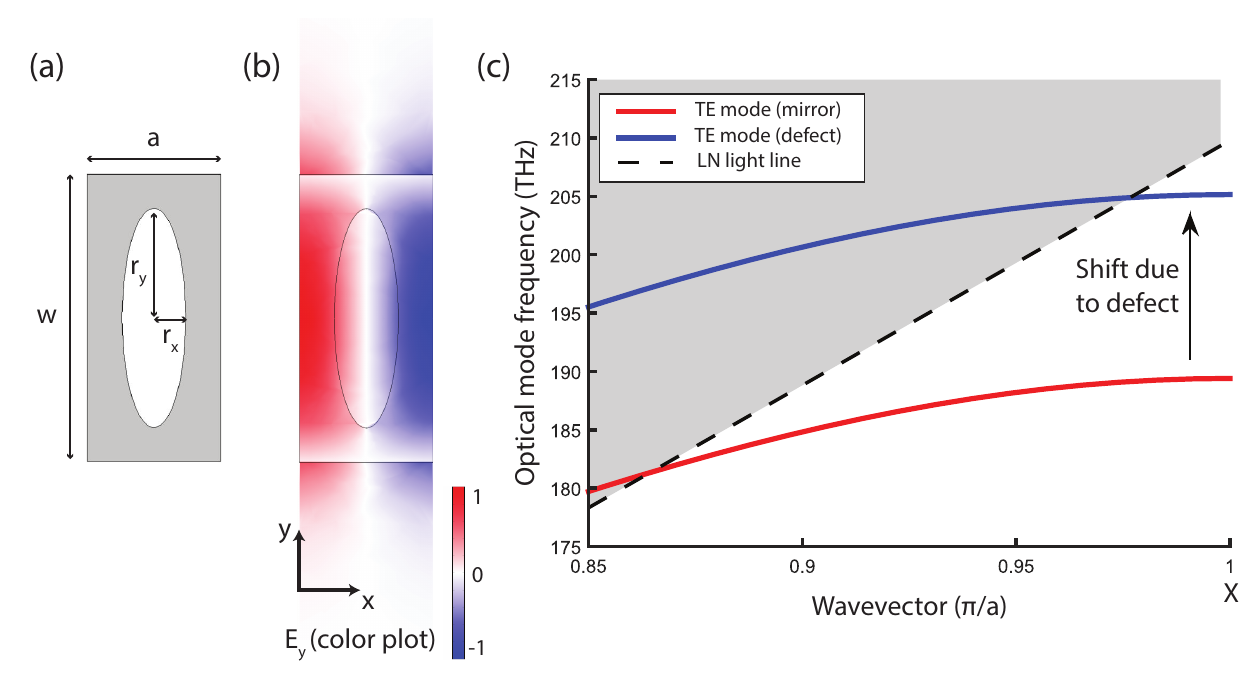}
\caption{(a) Unit cell geometry for the cavity mirror region.  The design parameters are: $a = 325~\text{nm}$, $w = 630~\text{nm}$, $r_x = 70~\text{nm}$, $r_y = 240~\text{nm}$. The Si device layer thickness is 220 nm.  (b) Plot of electric field y-component for the $X$-point mode of the nanobeam unit cell, with the cell geometry outlined in black. (c) Band diagrams showing the TE-like dielectric modes for the nanobeam mirror and defect regions.  The defect here is a 10\% reduction in the photonic crystal lattice spacing. The defect mode is chosen to lie near the LN light line, but still within the TE band gap of the mirror region.  Note that the TM-like modes are all above the LN light line. }
\label{fig:band_diagram}
\end{center}
\end{figure}

Given the constraints  described above, the mirror region unit cell shown in Fig. \ref{fig:band_diagram}(a) is designed to strike a balance between confining the light within the photonic crystal resonator while simultaneously having sufficient overlap with the LN substrate to take advantage of LN's nonlinear and electro-optic properties.  The $X$-point dielectric mode for the unit cell is shown in Fig. \ref{fig:band_diagram}(b), with the electric field $y$-component shown in a color scale.  For this unit cell, approximately 15\% of the electromagnetic energy is contained in the LN.  The TE-like dielectric band for this nominal unit cell is shown as the red curve in Fig. \ref{fig:band_diagram}(c).  Notice that for this design, the TE-like air band as well as all TM-like modes are above the LN light line, and are therefore excluded from the diagram.

\begin{figure}[h]
\begin{center}
\includegraphics[width=5in]{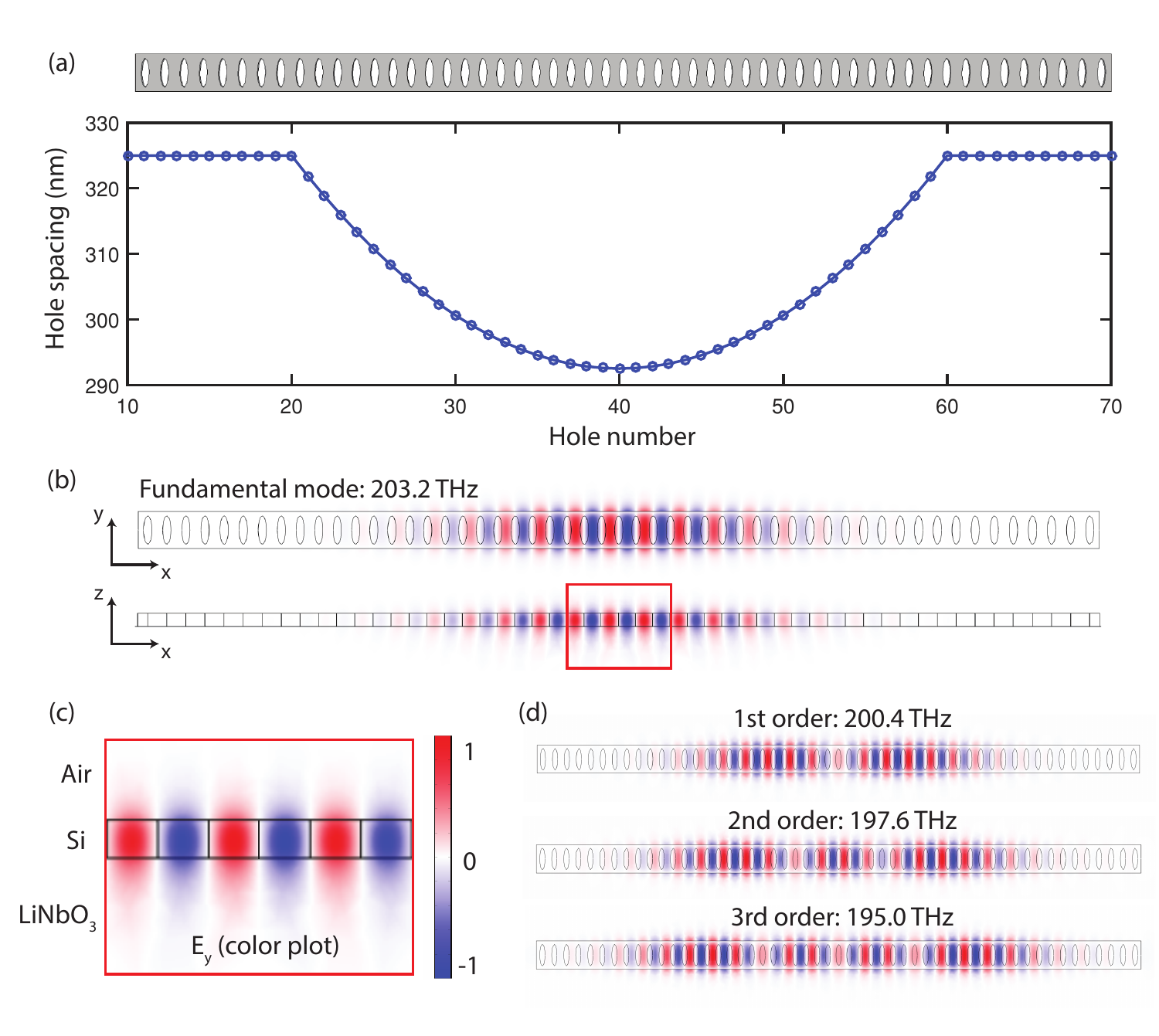}
\caption{(a) The unit cell length (ie. the lattice spacing) vs. hole number along the length of the nanobeam.  The nanobeam has a 39-hole defect consisting of a quadratic reduction in lattice spacing, down to a minimum of 90\% of the nominal spacing.  On either side of the defect are mirror regions each consisting of 20 unit cells with the nominal spacing of 325 nm. Not all mirror holes are shown.  (b) shows side and top views of the fundamental optical mode, which has a frequency of 203 THz.  The color plot shows the y-component of the optical mode electric field.  For both side and top views the cross-sections are taken through the center of the nanobeam.  The black lines show the outline of the device (the vertical lines in the side view mark the ellipse centers). (c) shows an enlarged image of the optical mode, showing the field penetration into the LN substrate.  Approximately 15\% of the electromagnetic energy is contained in the LN. (d) The nanobeam cavity supports various higher order longitudinal modes, separated by 2.6 to 2.8 THz.}
\label{fig:defect}
\end{center}
\end{figure}

From here, we move to designing the defect region. For nanobeam resonators, a defect with a quadratic profile has been shown to produce optical modes with smooth, Gaussian envelopes, resulting in low radiation losses and high quality factors \cite{Chan2009,Davanco2012, Li2015}.   We use a quadratic defect in which the hole lattice spacing is reduced by 10\%, while the size and shape of the elliptical holes are kept fixed.  The shift in the TE dielectric band due to this 10\% reduction is shown in Fig. \ref{fig:band_diagram}(c) (blue).  Cavities based on other types of defects (eg. varying the hole dimensions or beam width) are also possible. A typical beam design simulated in this paper included 39 defect holes in the center of the beam, surrounded on either side by mirror regions containing 20 unit cells each.  The hole spacing along the length of this nanobeam is shown in Fig. \ref{fig:defect}(a).

In order to judge the effectiveness of the cavity design, the optical modes of the structure were calculated using an electromagnetic finite-element solver (COMSOL).  There are several different sources of loss for  photonic crystal resonators, such as material absorption, scattering due to fabrication defects, as well as the leakage into radiation modes due to spatial confinement of the resonance.  In our simulations, we neglected the first two loss mechanisms (which typically set an upper bound on the measured $Q>5\times 10^6$), and considered  only the third, which is set fundamentally by the geometry and Maxwell's equations.
The radiation $Q$ factor, $Q_\text{rad}$, was determined by surrounding the simulation space with an absorbing layer so that far-field radiated light results in an imaginary eigenfrequency component.

\begin{figure}[h]
\begin{center}
\includegraphics[width=5in]{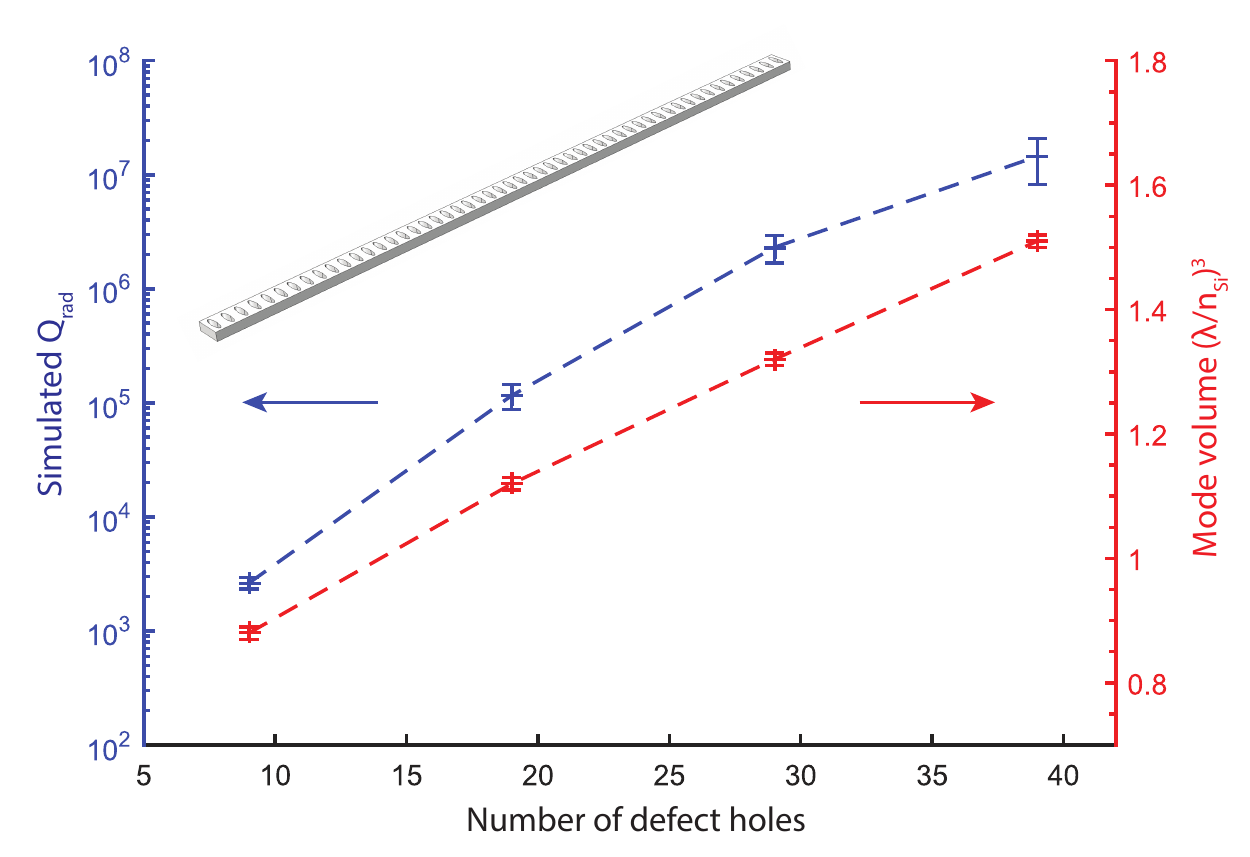}
\caption{Plot showing how the radiation-limited $Q$ factor (left axis) and mode volume (right axis) of the fundamental nanobeam mode changes with an increase in the number of holes in the defect region.  A longer defect region results in a larger mode volume, but also a greatly reduced amount of out-of-plane scattering.  The error bars for the mode volume and Q factor were established by varying the size of the simulation space and refining the mesh.  It should be noted that these finite-element simulations only consider losses due to far-field radiation from a perfect structure; in reality, measured device $Q$ factors will likely be limited by material absorption and fabrication defects \cite{Quan2010}. Inset: The geometry of the high-Q nanobeam with 39 defect holes. }
\label{fig:Q_vs_Length}
\end{center}
\end{figure}

One of the key ways to reduce radiation losses is to ensure that the transition from the mirror region to cavity defect occurs smoothly \cite{Akahane2003a}.  In the case of a quadratic defect with a fixed ``depth'', the defect becomes more adiabatic as the number of unit cells in the defect region is increased.  Figure \ref{fig:Q_vs_Length} shows the trend of increasing $Q_\text{rad}$ as the number of defect unit cells is increased.  At a defect size of 39 unit cells, the simulated $Q_\text{rad}$ exceeds 14 million.  The mode volume, important for understanding the quantum operation of the device, is also shown in Fig. \ref{fig:Q_vs_Length}.  As the defect size increases, the mode volume increases roughly linearly, up to about 1.5 $(\lambda/n_{Si})^3$. Here we use the standard definition of mode volume: $V = \frac{\int \epsilon |\vec{E}|^2 d^3r}{\text{max}\left( \epsilon |\vec{E}|^2\right)}$ \cite{Vuckovic2002}.
The error bars in Fig. \ref{fig:Q_vs_Length} are approximate and were established by varying the size of the simulation space and progressively refining the mesh used in the finite-element calculation.  For the last data point (39 defect unit cells) the estimate of the $Q_\text{rad}$ is limited by the computational size and mesh density rather than the cavity geometry.  As such, the value of 14 million for the ultimate $Q_\text{rad}$ should be taken as a lower bound. The material parameters used for all simulations in this paper are summarized in Table \ref{tab:parameters}.  It should be noted that the fraction of electromagnetic energy in the LN decreased modestly as the number of defect holes was increased, from 17\% at 9 defect holes to 13\% at 39 defect holes.  This is due to a change in the overall silicon filling fraction seen by the optical mode.

\begin{table}[b]
\centering
\caption{Material parameters used for simulations. }
\begin{tabular}{|c|c|c|c|}
\hline
Parameter & Description & Value & Ref.\\ \hline
$n_{Si}$  & Silicon refractive index (194 THz) & 3.48 &  \cite{Weber}\\
- & LN-orientation & X-cut & - \\
$n_o$  & LN ordinary refractive index (194 THz) & 2.21 & \cite{Nikogosyan2005} \\
$n_e$  & LN extraordinary refractive index (194 THz) & 2.14  & \cite{Nikogosyan2005}\\
$\epsilon_{11}$ & LN relative permittivity (DC) & 46.5 & \cite{Weis1985}\\
$\epsilon_{33}$ & LN relative permittivity (DC) & 27.3 & \cite{Nikogosyan2005} \\
$r_{13}$ & LN electro-optic coefficient & 9 pm/V & \cite{Weis1985}\\
$r_{33}$ & LN electro-optic coefficient & 31 pm/V & \cite{Weis1985}\\
$\epsilon'_{\text{Al}}$ & Aluminum relative permittivity (real part, 203 THz) & $-$208.2 & \cite{Rakic1998}\\
$\epsilon''_{\text{Al}}$ & Aluminum relative permittivity (imag. part, 203 THz) & 41.89 & \cite{Rakic1998}\\

\hline
\end{tabular}

\label{tab:parameters}
\end{table}

\section{Electro-optic Coupling}\label{sec:eo}

\begin{figure}[t]
\begin{center}
\includegraphics[width=\columnwidth]{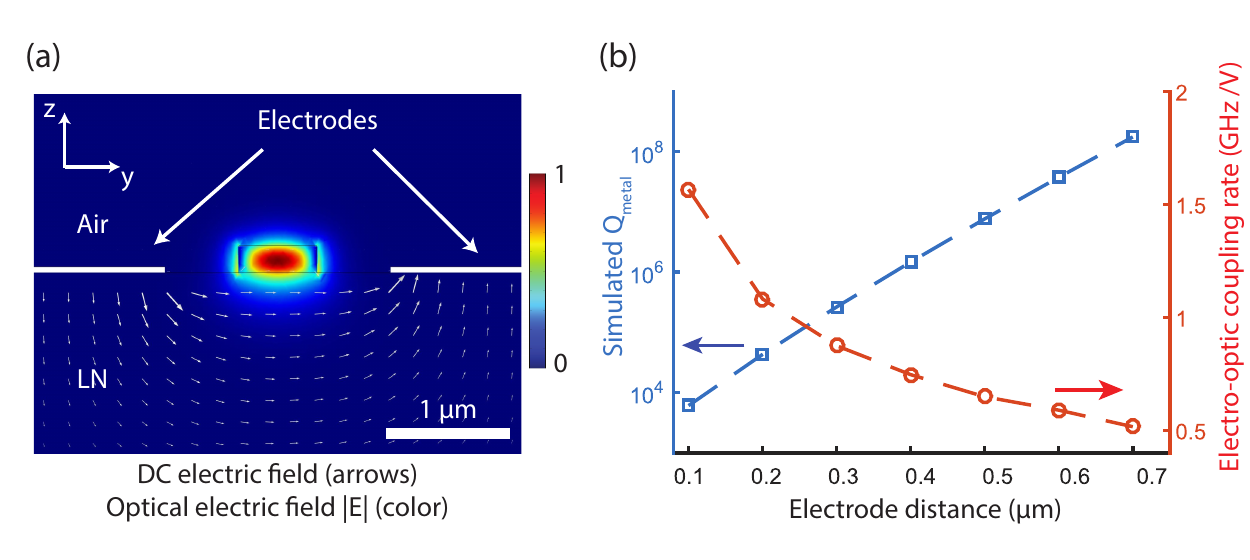}
\caption{(a) Plot showing the electro-optic overlap for the nanobeam cross-section.  The optical mode electric field norm is plotted in color, and the white arrows indicate the DC electric displacement field due to the electrodes. The nanobeam geometry is outlined in black and the electrodes are shown in white.  The distance between the electrodes and nanobeam edge is 600 nm and the electrode height is 50 nm. (b) shows how the simulated $Q_\text{metal}$ (due to metal absorption) and the electro-optic coupling rate $g_V/2\pi$ vary as a function of the electrode distance from the edges of the nanobeam.  }
\label{fig:Electro-optic}
\end{center}
\end{figure}


One of the key applications for photonic devices based on an LN platform is electro-optic modulation \cite{Chen2014,Chen2012,Chiles2014,Rao2015}. For a photonic crystal type cavity, electro-optic modulation can be achieved by fabricating electrodes near the cavity optical mode.  Applying a potential difference results in an electric field through the LN, which in turn causes a frequency shift of the optical mode via the electro-optic effect in LN. 

To judge the effectiveness of our nanobeam cavity design for electro-optic applications, we performed combined electromagnetic simulations to find the electro-optic coupling rate between the optical modes of a nanobeam defect unit cell and the DC to mmWave electric field generated by the electrodes.  In our simulation, we apply a fixed voltage $V_{\text{app}}$ between the two electrodes and calculate the resulting applied electric field $E^\text{app}$. This applied electric field in turn leads to an index perturbation $\Delta\varepsilon$ in the LN.  In LN the dielectric tensor is diagonal and the largest electro-optic coefficients are $r_{33}$ and $r_{13}$, so  $\Delta\varepsilon$ has three major components given by

\begin{equation}
\Delta \varepsilon_{xx(zz)}  = - r_{13} n_o^4 ~E^\text{app}_{y},
\end{equation}
and
\begin{equation}
\Delta \varepsilon_{yy}  = - r_{33} n_e^4 ~E^\text{app}_{y},
\end{equation}
where $n_o$ ($n_e$) is the ordinary (extraordinary) refractive index.  Notice that here we have taken the LN extraordinary crystal axis (often referred to as Z+) to lie along the y-axis of our simulation.

From first-order perturbation theory (see eg. \cite{Joannopoulos2008}), the frequency shift generated by an index perturbation $\Delta\varepsilon$ is given by 
\begin{equation}
\Delta\omega = -\frac{\omega_0}{2}~ \sum_{ij}~ \frac{\int_\textrm{LN} E^*_{0i} \Delta\varepsilon_{ij} E_{0j} ~d^3r}{\int E^*_{0i} \varepsilon_{ij} E_{0j} ~d^3r},
\end{equation}
where $\omega_0$ is the original resonance frequency, $E_0 = (E_{0x},E_{0y},E_{0z})$ is the electric field of the unperturbed optical mode, and the top integral is taken over the LN substrate region.  Finally, from this we can calculate the electro-optic coupling rate as
\begin{equation}
    \frac{g_V}{2\pi} = \frac{\Delta\omega}{2\pi V_{\text{app}}}.
\end{equation}

Figure \ref{fig:Electro-optic}(a) shows a cross-section of the nanobeam (cut along a dielectric segment), illustrating the overlap between the optical mode and the applied electric field from the electrodes.  Due to the high dielectric permittivity of LN at low frequencies, the applied electric field is almost completely confined to the LN substrate, increasing the overlap.

A key design parameter in such an electro-optic photonic device is the distance between the metal electrodes and the optical cavity.  Bringing the electrodes closer to the cavity results in a stronger electro-optic interaction per volt, but runs the risk of reducing the cavity $Q$ factor due to absorption. We define a loss parameter, $Q_\text{metal}$, to be the $Q$ due only to absorption losses in the metal electrodes.  For the simulations, we assumed the electrodes were aluminum (see Table \ref{tab:parameters}).
As the electrodes are moved farther from the nanobeam, the amount of field penetrating into the electrode region falls off exponentially, causing the $Q_\text{metal}$ (blue curve, left axis)  to increase dramatically.  However, the electro-optic coupling rate $g_V/2\pi$  (red curve, right axis) also decreases as the spacing is increased, suggesting  an optimal spacing that balances this trade-off.  It should be stressed that the $Q_\text{metal}$ reported here represents an approximate upper-bound on the real $Q$ factor which would include the radiation losses as well fabrication-induced and absorption losses.

A good design choice would be to choose an electrode distance where  $Q_\text{metal}$ is one or two orders of magnitude larger than the expected cavity quality factor.  For example, if we expect a fabrication-limited $Q$ of about $10^6$, then we can choose an electrode distance of 0.6 $\mu\text{m}$ (which has a simulated $Q_\text{metal} = 3.7\times 10^7$), and for this device we achieve a coupling rate of 0.59 GHz/V (4.3 pm/V). A more complete investigation of electro-optic modulation at GHz frequencies requires an analysis of the details of the driving circuit and will be studied in a future work. 

\section{Conclusions}

In summary, we have proposed a new way of confining light in wavelength-scale optical resonators on a hybrid Silicon/Lithium Niobate system. By using a thin-film bonded silicon top layer, we leverage many of the techniques of silicon photonics processing to design a cavity on LN.  We expect resonators of this type to have a wide range of applications, including achieving large coupling to isolated rare-earth ions (Er$^{3+}$:LiNbO$_3$) at telecom frequencies~\cite{mcauslan2009strong,zhong2015nanophotonic}, ultra-sensitive acousto- and electro-optic modulation, and achieving large three-wave mixing in resonant silicon devices.

\section*{Acknowledgements}   JW gratefully acknowledges support from a Stanford Graduate Fellowship. This work was supported by NSF ECCS-1509107 and the Stanford Terman Fellowship, as well as start-up funds from Stanford University. We thank Martin Fejer, Carsten Langrock, Jeff Chiles, Oskar Painter, and Johannes Fink for useful discussions.  Part of this work was performed at the Stanford Nano Shared Facilities (SNSF) and the Stanford Nanofabrication Facility (SNF).

\end{document}